\documentclass[twocolumn,showpacs,preprintnumbers,amsmath,amssymb,10pt]{revtex4}

\usepackage{graphicx}% Include figure files
\usepackage{dcolumn}% Align table columns on decimal point
\usepackage{bm}% bold math
\usepackage{color}

\newtheorem{process}{Process}
\begin{document}

%\preprint{$Revision: 1.102 $}

\title{Higher-order distributions and nongrowing complex networks without multiple connections}

\author{Tomas Hruz}
 \affiliation{Institute of Theoretical Computer Science, ETH Z\"{u}rich, Universit\"{a}tstrasse 6, 8092 Z\"{u}rich, Switzerland}
\author{Michal Natora}%
 \affiliation{Department of Software Engineering and Theoretical Computer Science, Berlin Institute of Technology, Strasse des 17 Juni 135,
10623 Berlin, Germany}
\author{Madhuresh Agrawal}
 \affiliation{Department of Computer Science and Engineering, Indian Institute of Technology Kanpur, Kanpur 208016, Utter Pradesh, India}

\date{\today}

\begin{abstract}

We study stochastic processes that generate non-growing complex networks
without self-loops and multiple edges (simple graphs). The work concentrates on
understanding and formulation of constraints which keep the rewiring stochastic
processes within the class of simple graphs. To formulate these constraints a
new concept of wedge distribution (paths of length 2) is introduced and its
relation to degree-degree correlation is studied. The analysis shows that the
constraints, together with edge selection rules, do not even allow to formulate
a closed master equation in the general case. We also introduce a particular
stochastic process which does not contain edge selection rules, but which, we
believe, can provide some insight into the complexities of simple graphs.

\end{abstract}

\maketitle

{\small
\section{Introduction}

Growth processes account for some phenomena observed in real networks and they
have been studied in many different scenarios (for survey see \cite{surveynew,
barabasi:survey, Evo}). Recently, more complex growth phenomena with node
addition and/or deletion are studied in \cite{aggr}, \cite{adddel} and
\cite{adddel2}. However, there are also many examples of networks, mainly in
biology (see \cite{Park} and associated references), in which a complex - and
not classically random  as in \cite{erdos} - topology is emerging but no growth
process occurs.  Instead of growth, one can observe a self-organizing
stochastic process which moves some network connections to different nodes.
Connections in these networks can also be dropped or created but the number of
edges stays statistically bounded within a given interval. Technological
networks like the Internet, once saturated in terms of size, also continue to
undergo stochastic changes in structure due to various types of reorganization.
Such processes, generally referred to as "non-growing" or "equilibrium
networks", constitute an important class of complex network dynamics and have
been studied within the framework of network statistical mechanics in
\cite{Evo} and \cite{japan}.

Although the dynamic behavior of non-growing networks has been studied in some
detail, one aspect remains elusive: to date, the general theory allows multiple
edges between nodes and self-loops connecting nodes to itself. Yet in most real
networks there are no instances in which self-loops or multiple edges would
provide a meaningful model of existing phenomena. It would therefore be
beneficial to create models that stay within the class of simple graphs (no
self-loops and multiple edges). For example in \cite{Evo}, a large collection
of real networks and their parameters is summarized, however in none of them
 multiple edges or self-loops would constitute a reasonable modeling feature.
Even in the cases where multiple edges do exist (as for example in Internet
routing network with backup links between the routers) it would be more useful
to introduce an edge capacity as a model than to introduce multiple edges. The
experience from modeling biological metabolic networks also shows that
biologists are more interested in being able to quantify the amounts of the
metabolites through an edge capacity than in a possibility of having more edges
between the nodes. Therefore, if more accurate models of real complex networks
are needed we have to understand how to model the constraints which would keep
the network without multiple edges and self-loops.

So far the methods used to approach this open problem \cite{mayerdorog} have
had three basic thrusts:

\noindent{\bf (i)} Under certain conditions, asymptotic behavior in size and in
time produces networks that are virtually free of self-loops and multiple
connections \cite{Evo} (i.e. very large networks will not develop many
self-loops and multiple edges). However, in the case of medium size networks
which do not grow over time, or in which the stochastic process persists for
extremely long time, this principle no longer applies in the same way and would
not explain the structural complexity characteristic of this class of simple
graphs. \\ \noindent{\bf (ii)} Specific constraints are applied to the processes
and initial conditions as in \cite{Park, merg}. \\ \noindent{\bf (iii)} Steady
state asymptotic features of processes for simple graphs are studied as in
\cite{mayerdorog}. This approach provides an insight into the asymptotic
behavior but if the states outside of an equilibrium are involved (transient
phenomena in time domain) or the network is not sufficiently large, a more
detailed description is needed.

In this article we introduce another line of thinking, where we study how the
constraints, which keep the network without multiple edges and self-loops can
be analytically formulated. Our arguments fall into the following parts:

\noindent{\bf (i)} The work concentrates on a basic process (hereinafter Simple
Edge Selection Process or S-ESP) in which an edge is randomly chosen and
rewired to a preferentially chosen vertex. The process is well established and
was studied in the network community (e.g. \cite{Evo}). It seems plausible to
take this process as a basic building element of a  non-growing network theory
as it is the simplest imaginable process that goes beyond classical random
graphs in terms of the process complexity. However, the S-ESP process does not
ensure that the graph remains within the class of simple graphs. To overcome
this modeling problem a simple modification is introduced, that precludes
formation of multiple edges and self-loops. The modified process is denoted
with SG-ESP (Simple Graph Edge Selection Process).
\\ \noindent{\bf (ii)} We study how to describe analytically the constraints
which preserve the simple graph property for the SG-ESP process. To express
these constraints more complex distributions than the degree distribution
$P(k)$ are needed. Namely, we introduce a new quantity $P(k,k',k'')$ called
wedge distribution, which describes the distribution of wedges (paths of length
2, see also Figure~\ref{fig:wedgeex}), where the end nodes have degree $k, k''$
and the middle node has degree $k'$. To be able to formulate the master
equation of the SG-ESP process it is also necessary to study in detail the
relations between $P(k)$, $P(k,k')$ and $P(k,k',k'')$, where $P(k,k')$ provides
information how the edges with degree $k$ on one side and the degree $k'$ on
the other side are distributed. The quantity $P(k,k')$ is often named
"degree-degree correlation", however as the analysis shows, to understand the
SG-ESP process we would also need distributions of more complex objects than
edges, and in that case a terminology using "degree-degree" terms would not be
of advantage. Therefore, we stick to the terminology where the distributions
are named according to the objects they describe.
\\ \noindent{\bf (iii)} Usage of the edge distribution $P(k,k')$, the wedge
distribution $P(k,k',k'')$ and the relations between them allows to formulate
the master equation for SG-ESP. A master equation is a phenomenological
equation that describes the dynamics of a certain quantity in a stochastic
process. In the case of complex networks, the basic quantity described with
master equations is the degree distribution. However, for the SG-ESP process
the equation is not closed i.e. contains quantities that would need more
equations to be sufficiently determined. The source of the problems is the edge
selection rule together with simple graph constraints. To gain an insight into
this complexity we propose to study a certain class of approximations to the
SG-ESP process which replaces edge selection with vertex selection rules and
the edge-rewiring rules with the deletion and creation of edges. As a concrete
instance of this class we define a process called VADE (Vertex Based Addition
and Deletion of Edges) and provide experimental evidence that it can
approximate SG-ESP in a certain parameter range. This approximation by no means
provides a solution to the problem of simple graph constraints, but it shows
one possible direction how to understand more about the structure of the SG-ESP
process. It is also interesting that the degree distribution master equation is
not enough to describe the VADE process, but instead an equation for the edge
distribution fully captures its behavior. In that sense VADE lies between
classical random graphs and the SG-ESP process.
\\ \noindent{\bf (iv)} We introduce a systematic method to derive master
equations for complex processes, where many different configuration cases can
make the derivation error prone or very difficult. The method allows us to
derive the master equation for the VADE process in a very efficient way.
However, even if the equation is solvable in principle, its nonlinear character
and complexity would need new methods to provide an insight into its solutions.

The paper is organized as follows: we start with an analysis of a well known
S-ESP process \cite{Evo}; then we study joint distributions of edges and
wedges; this leads us to section~\ref{sec:rate-swap} where we develop a master
equation for the SG-ESP process. As an attempt to provide a deeper
understanding into the structure of SG-ESP we introduce an approximation
process called VADE in section~\ref{sec:vade}. The VADE process is simpler than
SG-ESP as its master equation shows, and the simulation results suggest that
for some parameter ranges it can approximate SG-ESP very well.

\section{\label{sec:SG-ESP}Network Evolution with Simple Graph Constraints}

Suppose as a starting point a simple graph $G(V,E)$ with edge set $E$, $|E|=L$
and vertex set $V$, $|V|=N$. We denote with $N(k)$ the number of nodes
having degree $k$, and then degree distribution denoted as $P(k)$ can be
expressed as $P(k)=N(k)/N$. The averaging of a quantity $X$ is denoted with
$\langle X\rangle$ or with $\bar X$. Specifically, the average degree of the
network is denoted with $\bar k$, and it equals $\bar k=2L/N$. In the cases
where we consider changes of the basic quantities like $P(k)$ or $N(k)$ over
time, we add the parameter $t$ as in $N(k,t)$ or $P(k,t)$.

In the next section we study the S-ESP process and we introduce a systematic
method how to derive a master equation for processes like S-ESP. This method is
also used in the following sections to derive master equations for more complex
processes.

\begin{figure} \centering \includegraphics[width=7cm, height=9cm,
keepaspectratio]{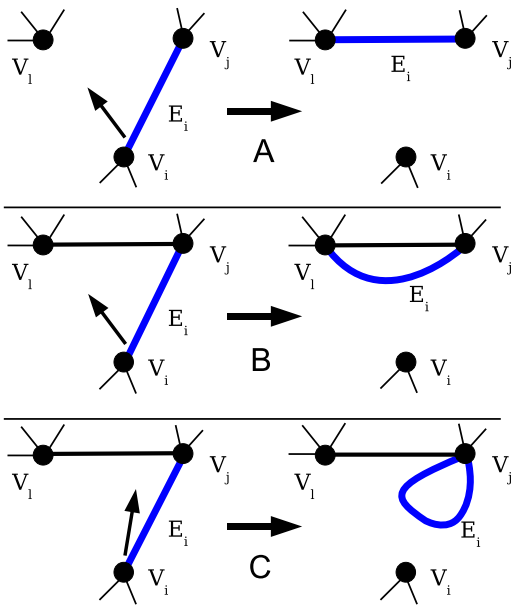} \caption{\label{fig:s-esp}(color online)
Illustration of the S-ESP process. The process selects a random edge and
rewires it to a node which was preferentially selected. A detailed explanation
of preferential selection and the process definition is provided in
Section~\ref{sec:SG-ESP}. The rewiring is unconditional, this means that there
is no test of the connectivity state between $V_l$ and $V_j$. During the
rewiring operation the following three cases can occur: In the case A there is no
edge between the node $V_l$ and the node $V_j$, therefore the process does not
create a multiple edge or a self loop and it will keep the simple graph
property. In the case B there is already an edge between$V_l$ and $V_j$, and
the process adds another edge between the nodes. In the last case C, if $V_l =
V_j$ the process creates a self-loop.} \end{figure}

\subsection{Simple Edge Selection Process}

The Simple Edge Selection Process (S-ESP) is defined by the following steps
that are supposed to be repeated in each discrete time unit $t$ (see
Fig.~\ref{fig:s-esp}):

\begin{process}[S-ESP] \hspace{1cm}

\begin{enumerate}

\item An edge, called $E_i$, is selected uniformly at random.

\item An end vertex, called $V_i$, of $E_i$ is selected uniformly at random. The
other end vertex will be called $V_j$.

\item \label{S-ESP:step:pref} A vertex, called $V_l$, is selected with a
probability proportional to $f(k)$ i.e. with probability $\frac{f(k)}{N\langle
f\rangle}$ where $k$ is the degree of $V_l$ and $\langle f\rangle$ is the mean
value of $f$, $\langle f\rangle \equiv \sum\limits_{s}f(s)P(s)$.

\item The edge $E_i$ is rewired from $V_i$ to $V_l$ i.e. the edge $E_i$ between
$V_i$ and $V_j)$ is deleted and a new edge between $V_l$ and $V_j$ is created.

\end{enumerate}

\end{process}

The process keeps the number of nodes and edges in the network constant, and
the function $f(k)$ models preferential attachment. Preferential attachment is
a concept studied in much detail (see for example \cite{barabasi:survey}) that
aims at modeling how the agents (processes) in complex networks would
choose/search an object on which they want to operate. For example in the
complex network of web pages and links, if a web page is updated, the author is
often providing links (edges) to other web pages. The links are going to pages
(nodes) which are known to the author and these are the pages having already
many links pointing to them. In this case, the choice where to link is modeled
choosing the function $f(k)$ as $f(k)=k$, i.e. the nodes in
step~\ref{S-ESP:step:pref} of the S-ESP process are selected with higher
probability if they have higher degree.

The S-ESP process captures some aspects, like preferential attachment, of
processes in real networks, however, this process generates multiple
connections and self-loops, which are rarely observed in real networks.
Multiple connections are created, because the edge is rewired irrespective of
already existing connections between $V_j$ and $V_l$. On the other hand,
self-loops emerge when vertex $V_l$ is the same vertex as $V_j$.

To derive the master equation describing the dynamics of the degree distribution for
the process above we start with a derivation of the probability term expressing
the probability that $V_i$ is of degree $k'$ and that $V_l$ is of degree
$k'''$. The master equation can be constructed parameterizing this term with
various values of $k'$ and $k'''$ which express all possible changes in degree
of vertices touched by the process step.

Selecting uniformly at random an edge and then selecting uniformly at random an
end vertex of this edge is equal to selecting a vertex linearly preferentially
\cite{Evo} i.e. with preference function $f(k)=k$. Therefore, the probability
that vertex $V_i$ has degree $k'$ is given by

\begin{equation}\label{}
\text{\bf Prob[}deg(V_i)=k'\text{\bf ]}=\frac{k'}{N\bar k}N(k')
\end{equation}

Analogically, the probability that $V_l$ is of degree $k'''$ is given by
$\frac{f(k''')}{N \langle f\rangle}N(k''')$. Since these two selections are
independent of each other  and since the order of selection is strictly
determined, the probability that $V_i$ is of degree $k'$ and that $V_l$ is of
degree $k'''$ is

\begin{eqnarray}\label{BT:S-ESP} \text{\bf Prob[}deg(V_i)=k'\text{ and
}deg(V_l)=k'''\text{\bf ]} && \nonumber\\ = \frac{k'}{N\bar
k}N(k')\frac{f(k''')}{N \langle f\rangle}N(k'''). \end{eqnarray}

To compute the number $N(k,t+1)$ of vertices having degree $k$ at time $t+1$
one has to consider the number of vertices $N(k,t)$ having degree $k$ already
at time $t$ minus the number of all vertices that changed their degree from $k$
to any other value $k' \neq k$, plus the number of all vertices that changed
their degree from any value $k' \neq k$ to $k$. A list of all cases in which
the quantity $N(k,t)$ changes its value is provided in Table~\ref{tab:S-ESP}.
For example, the first row in Table~\ref{tab:S-ESP} means that if vertex $V_i$
had degree $k+1$ and vertex $V_l$ had degree $k-1$ at time $t$, both vertices
will have degree $k$ at time $t+1$. Thus, compared to the situation at time $t$
there are two vertices more with degree $k$ at time $t+1$ and this is expressed
in the third column of the table.

\begin{table}
\centering
\begin{tabular}{c|c|c}\firsthline
Degree of $V_i$ & Degree of $V_l$ & Change in $N(k)$ \\\hline\hline
$k+1$ & $k-1$ & $+2$\\
$k$ & $k$ & $-2$\\
$k+1$& $k''' \neq \{k-1,k \}$& $+1$\\
$k' \neq \{k+1,k \}$ &$k-1$& $+1$\\
$k$& $k''' \neq \{k-1,k \}$& $-1$\\
$k'\neq \{k+1,k \}$ &$k$ &$-1$\\\hline
\end{tabular}
\caption{\label{tab:S-ESP} The table provides a list of all cases at time
$t$ that lead to a change in the quantity $N(k,t)$ at time $t+1$. The minimal
configuration describing changes in $N(k,t)$ for the S-ESP process is shown in
Figure~\ref{fig:s-esp}. The probability that the configuration changes is equal
to $\frac{k'}{N\bar k}N(k')\frac{f(k''')}{N \langle f\rangle}N(k''')$ as
described in the text. The table is used to derive the master equation for the
S-ESP (and SG-ESP) process. The master equation is generated using every row in
the table exactly once to parameterize the probability configuration term above
which describes the probability that the quantity N(k,t) will change. If the
degree parameter column contains a set of degree values (as for example in the
third row, second column) a summation over this set must be used. The third
column contains a multiplier expressing how much the quantity $N(k,t)$ would
change.}
\end{table}

Now every row of Table~\ref{tab:S-ESP} is used to generate one term in the
master equation as a parametrized version of the basic probability
term~(\ref{BT:S-ESP}). If the degree parameters in the first two columns
specify a set of degree values $k$, then a summation over this set must be
used. After having calculated the probability of each case multiplied by its
effect (the third row of Table~\ref{tab:S-ESP}), we can write down the master
equation for the number of vertices with degree $k$ at time $t+1$:

\begin{eqnarray}
N(k,t+1) = N(k,t) && \nonumber\\ + 2\times\frac{k+1}{N \bar k}N(k+1,t)
\frac{f(k-1)}{N \langle f \rangle }N(k-1,t)\nonumber&&\\ - 2\times\frac{k}{N
\bar k}N(k,t) \frac{f(k)}{N \langle f \rangle }N(k,t)\nonumber\\ +
1\times\sum\limits_{k''',k'''\neq \{k-1,k \}} \frac{k+1}{N \bar k}N(k+1,t)
\frac{f(k''')}{N \langle f \rangle }N(k''',t)\nonumber&&\\ + ...&&
\end{eqnarray}

Writing down all cases and using the fact that $\sum\limits_{s}f(s)N(s) =
\langle f \rangle N$ the master equation of the S-ESP process \cite{Evo} is
obtained:

\begin{eqnarray}\label{eq:S-ESP}
N(k,t+1) = N(k,t)  - \frac{f(k)}{N \langle f \rangle}  N(k,t)\nonumber&&\\
+ \frac{f(k-1)}{N \langle f \rangle}  N(k-1,t) - \frac{k}{N\bar k} N(k,t) +
\frac{k+1}{N\bar k}  N(k+1,t)
\end{eqnarray}
Note that in the above derivation the probability that one selects twice the same vertex, i.e. $V_l=V_i$, is not taken into account, since this probability is small. Nevertheless, the resulting master equation \ref{eq:S-ESP} is not fully correct.  This issue is addressed in \cite{evans:bipartite}. 

\begin{figure} \centering \includegraphics[width=7cm, height=9cm,
keepaspectratio]{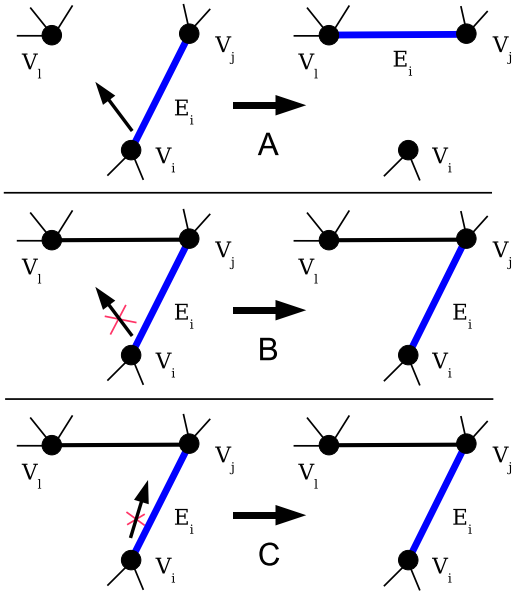} \caption{\label{fig:sg-esp}(color online)
Illustration of the SG-ESP process. The process selects a random edge and
rewires it to a node which was preferentially selected. A detailed explanation
of preferential selection and the process definition is provided in
Section~\ref{sec:SG-ESP}. In contrast to S-ESP, according to the SG-ESP process
the rewiring occurs only if $V_l \neq V_j$ and if $V_l$ is not directly
connected by an edge to $V_j$. The process verifies first if there is an edge
between $V_l$ and $V_j$ and if there is no one (case A) the rewiring takes
place, otherwise (cases B and C) the process leaves the original edge in
place.}
\end{figure}

\subsection{Simple Graph Edge Selection Process}

A natural extension of the S-ESP process, which would keep the process within
the class of simple graphs is to introduce a connectivity test whether there is
an edge between the target nodes, and if one is detected, the rewiring would
not take place. The process called Simple Graph Edge Selection Process (SG-ESP)
can be defined as follows (see also Fig. \ref{fig:sg-esp}):

\begin{process}[SG-ESP]

The following steps are repeated on a graph $G$ in each discrete time unit $t$.

\begin{enumerate}
\item \label{SG-ESP:step:edge}An edge, called $E_i$, is selected uniformly at
random.
\item An end vertex, called $V_i$, of $E_i$ is selected uniformly at random. The
other end vertex will be called $V_j$.
\item A vertex, called $V_l$, is selected with a probability proportional to
$f(k)$ as in step~\ref{S-ESP:step:pref} of the S-ESP process.
\begin{enumerate}\item \label{SG-ESP:step:test} Check if $V_l$ is an end vertex
of $E_i$ or if $V_l$ is directly connected to $V_j$. If so, skip the
next step. \end{enumerate}
\item The edge $E_i$ is rewired from $V_i$ to $V_l$.
\end{enumerate}

\end{process}

The only difference to the S-ESP process is the step~\ref{SG-ESP:step:test}.
However, to express analytically what this means, one has to know when the
condition in step~\ref{SG-ESP:step:test} is (not) satisfied. This means to know
the probability that a vertex $V_l$ is (not) directly connected to an edge
$E_i$, in other words $V_l$ is (not) a direct neighbor of $V_j$.

One of the main goals of the present paper is to show that we need to
study not only edge distributions but also distributions of higher order
objects (wedges) to understand the SG-ESP process in the class of simple
graphs. In the next sections we investigate the relations of these
distributions and show that they are necessary to analytically express the
condition in step~\ref{SG-ESP:step:test} of the SG-ESP process.

\begin{figure}
\centering
\includegraphics[width=4cm, height=4cm, keepaspectratio]{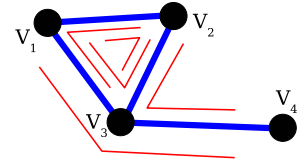}
\caption{\label{fig:wedgeex}(color online) Wedges are paths of length
2. The graph in this example contains the following 5 wedges: $(V_3,V_1,V_2)$,
$(V_1,V_2,V_3)$, $(V_2,V_3,V_1)$, $(V_1,V_3,V_4)$, $(V_2,V_3,V_4)$. The wedge
distribution $W(k,k',k'')$ is defined in section~\ref{subsec:higherorder}. In
this example the wedge distribution is $W(1,3,2)=2$, $W(2,3,2)=1$,
$W(2,2,3)=2$.}
\end{figure}

\subsection{\label{subsec:higherorder}Higher order distributions}

We use the term higher order distribution generally to denote a distribution of
more complex objects than nodes. Apart from the degree distribution $P(k)$, the
first higher order distribution, which is necessary for further analysis,
describes a distribution of edges and is defined as

\begin{align}
P(k,k') = \begin{cases} \frac{1}{2}\frac{L(k,k')}{L},&\text{if } k \neq k',
\\ \frac{L(k,k')}{L},&\text{if } k = k'. \end{cases}
\end{align}

where $L(k,k')$ denotes the number of edges in the network whose end vertices
have degree $k$ and $k'$, and $L$ denotes the total number of edges in the
network. The distribution $P(k,k')$ does not describe directly the
distribution $\frac{L(k,k')}{L}$ but there are two important reasons why the
above form of the edge distribution is necessary: a) $P(k,k')$ can be
interpreted as a probability that a vertex with degree $k$ has a neighbor with
the degree $k'$, and this form can be directly used for the analysis of the
SG-ESP process; b) it is technically simpler to express the relation
between $P(k,k')$ and $P(k)$ as between $\frac{L(k,k')}{L}$ and $P(k)$.
Moreover, if needed, $\frac{L(k,k')}{L}$ can be easily computed from $P(k,k')$.

\begin{figure}
\centering
\includegraphics[width=4cm, height=4cm, keepaspectratio]{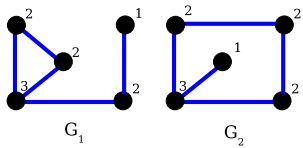}
\caption{\label{fig:edge-node-counter-example} (color online) Two
non-isomorphic graphs $G_1$ and $G_2$ with $N=5, L=5$, where the node degrees
are shown. The graphs have the same degree distribution $P(k)
=(P(1)=\frac{1}{5}, P(2)=\frac{3}{5}, P(3)=\frac{1}{5})$ but different edge
distributions $P_{G_1}(k,k')= (P(1,2)=\frac{1}{10}, P(1,3)=0,
P(2,2)=\frac{1}{5}, P(2,3)=\frac{3}{10})$ and $P_{G_2}(k,k')= (P(1,2)=0,
P(1,3)=\frac{1}{10}, P(2,2)=\frac{2}{5}, P(2,3)=\frac{2}{10})$. The example
shows that in general the degree distribution $P(k)$ can not uniquely determine
the edge distribution $P(k,k')$, but $P(k,k')$ determines $P(k)$ according to
equation \ref{eq:node-edge}.}
\end{figure}

The edge distribution $P(k,k')$ is symmetric, $P(k,k')$=$P(k',k)$, because in
the definition of $L(k,k')$ we cannot distinguish which end vertex of the edge
has degree $k$ resp. $k'$. It was reported \cite{clust}, that

\begin{equation} \label{eq:node-edge}
P(k) = \frac{\bar k}{k} \sum\limits_{k'}P(k,k')
\end{equation}

and that for uncorrelated networks the edge distribution factorizes, namely
\mbox{$P(k,k') = \tilde P_{L}(k)\tilde P_{L}(k')$} where \mbox{$\tilde P_{L}(k) = \frac{k
P(k)}{\bar k}$.}

The next component needed for analysis of the SG-ESP process is a distribution
describing probabilities of higher order objects called wedges. A wedge is an
object formed by three vertices which are connected together by two edges (a
path of length 2, see Fig. \ref{fig:wedgeex}). Similarly as for edges, we
define the wedge distribution as:

\begin{align}
P(k,k',k'') = \begin{cases} \frac{1}{2}\frac{W(k,k',k'')}{W},&\text{if } k \neq
k'', \\ \frac{W(k,k',k'')}{W},&\text{if } k = k''. \end{cases}
\end{align}

where $W(k,k',k'')$ is the number of wedges with the middle node having degree
$k'$ and one of the end nodes with degree $k$ and the other with the degree
$k''$. The number of all wedges in the network is denoted with $W$. The wedge
distribution is symmetric in both outer arguments, $P(k,k',k'') = P(k'',k',k)$,
but in general $P(k,k',k'') \neq P(k',k,k'')$ and $P(k,k',k'') \neq P(k,k'',k')$.

A vertex of degree $k$ is a middle vertex of $\binom{k}{2}= \frac{k^2}{2} -
\frac{k}{2}$ wedges. Therefore the total number of wedges is fully
determined by the degree distribution $P(k)$ and is given by $W =
\sqrt{2L^3C_{0}}$, where $C_{0}=\frac{ \bar k}{N} {\left( \frac{\langle k^2
\rangle - \bar k}{ \bar k^{2}}\right)}^2$. It can be shown \cite{clust}, that
$C_{0}$ equals to the clustering coefficient in the case of an uncorrelated
network.

To derive the relation between the edge and the wedge distribution we have to
consider that $k+k'-2$ wedges pass through an edge of degree $(k,k')$. Further,
considering how many edges a wedge can contribute to and using correct
normalization, the following relation for $k,k'>1$ is obtained:

\begin{equation}\label{eq:edge-wedge}
P(k,k') =  \sqrt{2LC_{0}}
\frac{\sum\limits_{k''}P(k,k',k'')+P(k'',k,k')}{k+k'-2}
\end{equation}

In a different context and differently defined, the distributions of more
complex objects have also been studied in \cite{wegde, subgraph, clustcomp}.
Another confirmation of the importance of higher order distributions comes from
the studies of complex interconnection patterns called network motifs. As the
authors of \cite{MotifsAsBlocks} reported, network motifs observed in the real
networks clearly distinguish these networks from artificial randomly generated
instances.

\begin{figure} \centering \includegraphics[width=4cm, height=4cm,
keepaspectratio]{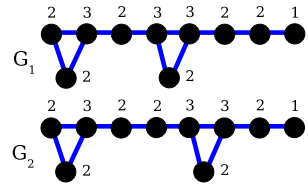}
\caption{\label{fig:wedge-edge-counter-example} Two non-isomorphic graphs $G_1$
and $G_2$ with $N=10, L=11$ and $W=15$, where the nodes are labeled with the
degrees. The graphs have the same edge distribution $P(k,k')=
(P(1,2)=\frac{1}{22}, P(2,2)=\frac{2}{11}, P(2,3)=\frac{7}{22},
P(3,3)=\frac{1}{11})$ but different wedge distributions $P_{G_1}(k,k',k'')=
(..., P(2,2,3)=\frac{1}{10}, ...)$ and $P_{G_2}(k,k',k'')= (...,
P(2,2,3)=\frac{2}{15}, ...)$. In general, an edge distribution can not
determine the wedge distribution of a graph, but $P(k,k',k'')$ determines
$P(k,k')$ according to equation \ref{eq:edge-wedge}.} \end{figure}

\subsection{Connection probabilities}

The wedge distribution introduced in the previous section can be used to
express various connectivity probabilities between nodes, edges and wedges. The
connection probabilities in turn allow to describe the simple graph
constraints.

Assume an arbitrary vertex of degree $k$ and an arbitrary edge of degree
$k',k''$. If $k$ is different from $k'$ and $k''$, the vertex cannot be part of
an edge with degree $k',k''$. In this case, we can consider a bipartite
subgraph, where one partition is represented by $N(k)$ (all vertices of degree
$k$) and the other partition is represented by all vertices of the edge set
$L(k',k'')$. In this subgraph there are $N(k)L(k',k'')$ possible edges, but
only $W(k,k',k'')$ of them are present. Therefore, the probability that an
arbitrary vertex of degree $k$ and an arbitrary edge of degree $k',k''$ are
part of the same wedge is $W(k,k',k'')/(N(k)L(k',k''))$. Similar considerations
can also be used for the cases $k=k'$ and $k=k''$, therefore the probability
$P_{k,(k',k'')}$ that a vertex of degree $k$ is directly connected to a vertex
of degree $k'$ which is a part of an edge of degree $k',k''$ is given by

\begin{equation}
\label{eq:connedge} P_{k,(k',k'')} = \frac{2W \cdot
P(k,k',k'')}{(N(k)-\delta_{k,k'}-\delta_{k,k''})L(k',k'')}
\end{equation}

where $\delta_{i,j}$ denotes the Kronecker delta. Analogous reasoning for
two distinct vertices leads to the probability that a vertex of degree $k$ is
directly connected to vertex of degree $k'$. This probability will be denoted
as $P_{k,k'}$ and is given by

\begin{equation}
\label{eq:conn} P_{k,k'} = \frac{2L \cdot
P(k,k')}{(N(k)-\delta_{k,k'})N(k')}.
\end{equation}

If $k \neq k'$ then the edges $L(k,k')$ form a bipartite subgraph with the
partitions $N(k)$ and $N(k')$. Therefore, an arbitrary vertex of degree $k$
participates in $L(k,k')/N(k)$ edges of degree $k,k'$. In general, the number
of edges of degree $k,k'$, which an arbitrary vertex of degree $k$ participates
in, is denoted as

\begin{equation} \label{eq:T}
T(k,k')=\bar k \frac{P(k,k')}{P(k)}.
\end{equation}

In an analogous way, the number of wedges of degree $k,k',k''$ which an
arbitrary edge of degree $k,k'$ participates in is

\begin{equation}
\label{eq:T2}
\sqrt{2LC_{0}}\frac{2P(k,k',k'')}{(2-\delta_{k,k'})P(k,k')}.
\end{equation}

This quantity can be also interpreted as the number of edges of degree $k',k''$
that share the vertex of degree $k'$ with another edge of degree $k,k'$.

The wedge distribution is useful in many other ways to express the
probabilities of connectivity resp. neighborhood relations. For example, the
probability that a vertex which is a nearest neighbor of an edge of degree
$k,k'$ has degree $k''$ is given by
\begin{equation*}
P(k''|k,k')=\frac{P(k'',k,k')}{\sum\limits_{k''}P(k'',k,k')}.
\end{equation*}

The wedge distribution can also be useful in obtaining more accurate
estimations of the local clustering coefficient compared to the situation when
only the edge distribution is used (see \cite{Natora2007} for further results concerning the wedge distribution). 

\subsection{\label{sec:rate-swap} Master equation for SG-ESP}

To derive the master equation of the SG-ESP process we follow the method
introduced at the beginning of this section for the S-ESP process. First, we
specify a general probability term describing, for the given configuration of
nodes (see Fig.~\ref{fig:sg-esp}), the probability that the configuration change
happens i.e. that the edge will be rewired. A list of all possible cases
leading to a change in the quantity $N(k,t)$ is the same as for S-ESP (see
Table~\ref{tab:S-ESP}).

Three probabilities have to be considered in order to evaluate the overall
probability of the configuration change. Firstly, as discussed in the previous
sections, $P(k',k'')$ is the probability that vertex $V_i$ in
Fig.~\ref{fig:sg-esp} is of degree $k'$ and it has a direct neighbor vertex
$V_j$ of degree $k''$. Secondly, $\frac{f(k''')}{N \langle f(k) \rangle
}(N(k''')-\delta_{k''',k'}-\delta_{k''',k''})$ is the probability that vertex
$V_l$, which is selected with a probability proportional to $f(k)$, is of
degree $k'''$ and is not a part of the edge $E_i$. The last probability to
consider is the probability that there is no edge between $V_l$ and $V_j$. It
is equal to $(1-P_{k''',(k'',k')})$ as discussed above. Putting all together,
the probability that the rewiring occurs is given by

\begin{eqnarray}
P(k',k'') \frac{f(k''')}{N \langle f(k) \rangle} (N(k''') -
\delta_{k''',k'}-\delta_{k''',k''}) \times&& \nonumber\\
(1-P_{k''',(k'',k')}).
\end{eqnarray}

The enumeration of all possible contributions to $N(k,t+1)$ is identical to
the S-ESP process, so after summarizing all of them, we obtain the following
master equation for the degree distribution. In the equation we suppose that
the quantities (as $N(k), P(k',k'')$ and $P_{k,(k',k'')}$) related to some
object distribution are functions of discrete time $t$, however in the
following text we omit the time variable $t$ to simplify the notation in cases
where its presence is clear from the context.

\begin{widetext}\begin{eqnarray}  \label{eq:swap} N(k,t+1) = N(k)  -
\sum\limits_{k',k''}P(k',k'')\frac{f(k)(N(k) - \delta_{k,k'}-
\delta_{k,k''})}{N \langle f \rangle }(1-P_{k,(k',k'')}) && \nonumber\\ +
\sum\limits_{k',k''}P(k',k'')\frac{f(k-1)(N(k-1) - \delta_{k-1,k'}-
\delta_{k-1,k''})}{N  \langle f \rangle}(1-P_{k-1,(k',k'')}) && \nonumber\\ -
\sum\limits_{k',k'''}P(k',k)\frac{f(k''')(N(k''') - \delta_{k''',k'}-
\delta_{k''',k})}{N \langle f \rangle}(1-P_{k''',(k',k)}) &&\nonumber\\+
\sum\limits_{k',k'''}P(k',k+1)\frac{f(k''')(N(k''') - \delta_{k''',k'}-
\delta_{k''',k+1})}{N  \langle f \rangle }(1-P_{k''',(k',k+1)}). \end{eqnarray}
\end{widetext}

To see the relation of this equation to the equation for S-ESP (\ref{eq:S-ESP})
all terms in the equation above can be expanded in a particular way. For
example, after expanding all factors of the last term on the right-hand side of
equation (\ref{eq:swap}), the term can be written as

\begin{eqnarray}
\label{eq:exp} \frac{k+1}{N \bar k}N(k+1) - && \nonumber \\
-\sum\limits_{k'}\frac{f(k')P(k',k+1)}{N\langle f \rangle} -
\frac{f(k+1)(k+1)}{N^2\langle f \rangle \bar k}N(k+1) - && \nonumber \\
-\sum\limits_{k',k'''}\frac{2W \cdot f(k''')P(k''',k',k+1)}{N\langle f \rangle
L(2-\delta_{k+1,k'})}
\end{eqnarray}

The other three terms of equation (\ref{eq:swap}) can be written in a similar
way. This form of equation (\ref{eq:swap}) shows that all terms of the S-ESP
process are contained in the master equation of the SG-ESP process (the first
term in expression (\ref{eq:exp}) corresponds to the last term in equation
(\ref{eq:S-ESP}) etc). The additional terms in (\ref{eq:swap}) arise from the
simple graph constraint and they show how well and under which conditions the
S-ESP process approximates the SG-ESP process. For example, the additional
terms become negligible in the limit of $N \gg 1, \frac{W}{N \cdot L}\ll 1$.
This is an example of conditions under which simple networks can be treated as
non-simple networks, but the equation (\ref{eq:swap}) also allows more complex
approximations to be investigated.

\begin{table}
\begin{tabular}{c|c|c|c}\firsthline
Graph & $P(k,k')$ & $P(k)$ & $\sum_{k'}k'P(k',2)$ \\\hline\hline
{} & {} & {} \\
\includegraphics[width=2cm, height=2cm,
keepaspectratio]{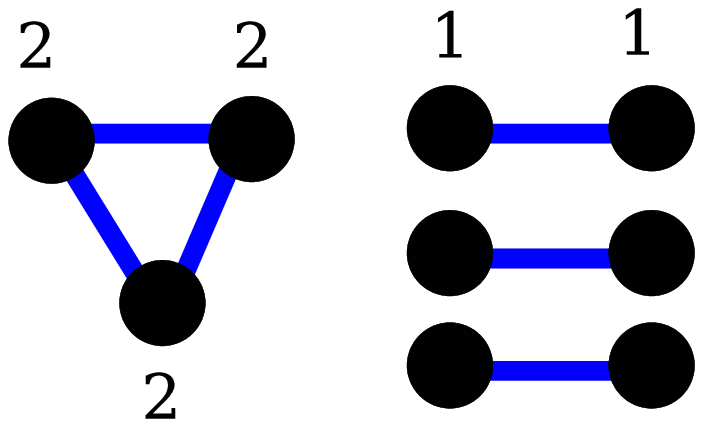} &$\begin{pmatrix} 1/2 & 0 \\ 0
& 1/2 \end{pmatrix}$ & $\begin{pmatrix} \bar k/2 & \bar k / 4 \end{pmatrix}$ &
1 \\
{} & {} & {} \\
\includegraphics[width=2cm, height=2cm,
keepaspectratio]{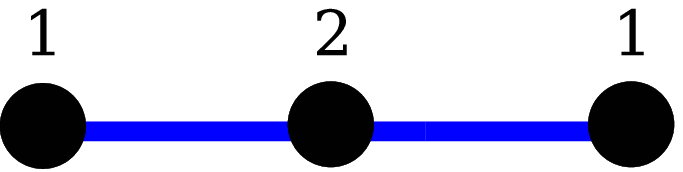} &$\begin{pmatrix} 0 & 1/2 \\
1/2 & 0 \end{pmatrix}$ & $\begin{pmatrix} \bar k/2 & \bar k / 4 \end{pmatrix}$
& 1/2 \\
\end{tabular}
\caption{This example shows that it is not possible to express
$\sum\limits_{k'}f(k')P(k',k)$ by $F(f(k),P(k))$, where $F$ denotes an
arbitrary function. Consider the two edge distributions in the second column of
the table, which describe edges with degree $k,k'=1,2$. Further assume that
$f(k) = k$. Both edge distribution lead to the same degree distribution (the
third column) according to equation \ref{eq:node-edge}, but the expression
$\sum_{k'}f(k')P(k',2)$ is different.}
\label{table:counter}
\end{table}

The counter-examples in
Figures~\ref{fig:edge-node-counter-example}~and~\ref{fig:wedge-edge-counter-example}
show, that in general it is not possible to obtain $P(k,k',k'')$ from $P(k,k')$
or $P(k,k')$ from $P(k)$. One can also show by counter-example (see table
\ref{table:counter}) that $\sum\limits_{k'}f(k')P(k',k)$ cannot be expressed
by $F(f(k),P(k))$, where $F$ denotes an arbitrary function. This means that
equation \ref{eq:swap} can not be written in a self-consistent way and hence is
not solvable in principle. To obtain a closed description we need to consider
adding equations for higher order distributions into the system.

\begin{center}
\begin{table*}
\begin{tabular}{c|c|c}\firsthline
Degree of $V_i$ & Degree of $V_j$ & Change in $L(k,k')$ \\\hline\hline
$k-1$ & $k-1$ & $+ 2T(k-1,k') $\\
$k-1$ & $\neq \{k,k',k-1,k'-1 \}$ & $+ T(k-1,k')$\\
$k'-1$& $k'-1$& $+2T(k'-1,k)$\\
$k-1 $ &$k'-1$& $ +T(k-1,k') + T(k'-1,k) + 1$\\
$k'-1$& $\neq \{k,k',k-1,k'-1 \}$& $ +T(k'-1,k)$\\
{}& {}& {}\\
$k$ &$k$ &$- 2T(k,k')$\\
$k$ &$\neq \{k,k',k-1,k'-1 \}$ &$- T(k,k')$\\
$k'$ &$k'$ &$- 2T(k',k)$\\
$k$ &$k'$ &$- T(k,k') - T(k',k)$\\
$k'$ &$\neq \{k,k',k-1,k'-1 \}$ &$ -T(k',k)$\\
{}& {}& {}\\
$k-1$ &$k$ &$+ T(k-1,k') - T(k,k')$\\
$k-1$ &$k'$ &$+ T(k-1,k') - T(k',k)$\\
$k'-1$ &$k$ &$+ T(k'-1,k) - T(k,k')$\\
$k'-1$ &$k'$ &$+ T(k'-1,k') - T(k',k)$\\
\hline

\end{tabular}
\caption{Similarly as in Table~\ref{tab:S-ESP} a list of the cases at time $t$
that will lead to a change in the quantity $L(k,k')$ at time $t+1$ is provided.
The table gives only the cases where an edge is added and where $k \neq k'$.
The cases where the edge is deleted are analogical. Cases that follow from
permutation are also not listed, for example, the case where $V_i$ has degree
$k'-1$ and $V_j$ has degree $k-1$ is not listed, because this case follows
directly from the case at line 4 by permutation. When considering all cases for
the edge addition (also the permuted ones) a total of 24 is found, analogically
we obtain 24 cases for the edge deletion.} \label{table:varVADE}
\end{table*}
\end{center}

\begin{figure} \centering \includegraphics[width=5.5cm, height=5.5cm,
keepaspectratio]{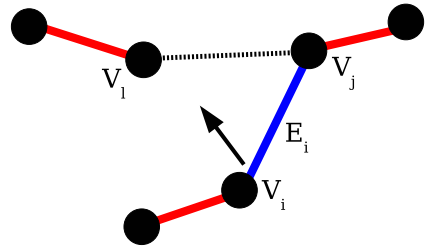} \caption{\label{fig:edgerate}(color online) The
randomly selected edge $E_i$ touches the edges (red colored) that share a
common end-vertex with $E_i$. Assuming that $V_i$ is of degree $k$ and $V_j$ is
of degree $k'$, the wedge distribution $P(k',k,k'')$ is needed in order to know
how many edges of degree $k,k''$ share $V_i$ as a common end-vertex with
$E_i$, see expression \ref{eq:T2}.} \end{figure}

To see what are the building blocks of SG-ESP master equations for higher order
distributions we consider first the edges. In order to derive a master equation
for the edge distribution, all situations at time $t$ which will contribute to
$L(k,k')$ at time $t+1$ must be enumerated. For example, if $V_i$ is of degree
$k+1$ we must know how many edges of degree $k+1,k'$ participate in $V_i$,
because all these edges will lose a connection and contribute to $L(k,k')$,
too. However, vertex $V_i$ was not selected independently (like vertex $V_l$),
but it is an end-vertex of the edge $E_i$ (see Fig. \ref{fig:edgerate}). Hence,
the number of edges of a certain degree which participate in vertex $V_i$ is
given by expression~\ref{eq:T2}. This expression is proportional to the wedge
distribution $P(k,k',k'')$. Therefore, the wedge distribution is needed to
derive the master equation for the edge distribution. Moreover, when trying to
derive master equations for wedges, the problem persists: In order to know how
many wedges of a certain degree share vertex $V_i$, which is again part of an
edge with prescribed degrees, we need the knowledge of a distribution which is
of a higher order than that for wedges.

The analysis above shows that to fully understand the behavior of the SG-ESP
process, a whole hierarchy of objects and their distributions must be involved.
How far this hierarchy above the wedge distribution is reaching would be very
difficult to know, however one conclusion is that the main source of the
problems is the edge selection Step~\ref{SG-ESP:step:edge} of the SG-ESP
process and the simple graph constraint. A possibility how to understand more about this complex phenomena and
still have a closed description of the process is to introduce more complex
vertex selection rules which would approximate the edge selection rule.

\section{\label{sec:vade} Vertex Based Addition and
Deletion of Edges}

In this section we describe a class of processes that preserves the simple
graph structure and that can be modeled by a single self-consistent equation.
This class will be called "VADE", which stands for "Vertex based Addition and
Deletion of Edges" and it aims at approximating the edge selection rule of
SG-ESP with vertex selection rules. The VADE process is defined as

\begin{process}[VADE] \hspace{1cm}

The following steps are repeated on a graph $G$ in each discrete time unit $t$.

\begin{enumerate}
\item With probability $q$ do the following:
\begin{enumerate}

\item \label{VADE:step:insert-start}choose a vertex $V_i$ with a probability
proportional to $f_1(k)$;

\item choose a vertex $V_j$ with a probability proportional to $f_2(k)$;

\item \label{VADE:step:insert-loop}if $V_i = V_j$ or if $V_i$ \emph{is}
directly connected to $V_j$, skip the next step;

\item add an edge between $V_i$ and $V_j$.

\end{enumerate}

\item With probability $1-q$ do the following:
\begin{enumerate}

\item \label{VADE:step:delete-start} choose a vertex $V_l$ with a probability
proportional to $g_1(k)$;

\item Choose a vertex $V_m$ with a probability proportional to $g_2(k)$;

\item \label{VADE:step:delete-loop} if $V_l = V_m$ or if $V_l$ is \emph{not}
directly connected to $V_m$, skip the next step;

\item delete the edge between $V_l$ and $V_m$.
\end{enumerate}
\end{enumerate}

\end{process}

The VADE process is similar to the SG-ESP process in several ways. It contains
preferential selection parameters ($f_1, f_2, g_1, g_2$), which give the
process the flexibility to approximate the edge selection rule of SG-ESP. If the
parameter $q$ is chosen appropriately as we discuss at the end of the
simulation section, the number of edges is stable in a narrow interval
approximating the constant number of edges in the SG-ESP process. The process
also preserves the simple graph structure, and the preferential selection
parameters can be set to generate a degree distribution which is similar to the
one generated by the SG-ESP process (see the simulation study in
Section~\ref{sec:simulations}).

The reason why the VADE process can be modeled by a self-consistent master
equation is that all the selection rules in the VADE process are defined for
vertices; in contrast to the SG-ESP process where an edge selection rule is
used. Namely, the probability that two vertices are directly connected with
each other requires the knowledge of $P(k,k')$ (see equation \ref{eq:conn}),
and also only from $P(k,k')$ we can derive how many edges of degree $k,k'$
participate in a vertex of degree $k$, see expression (\ref{eq:T}). This is the
reason why the dynamics of the VADE process is fully captured in a closed
master equation for the edge distribution.

To derive a master equation for this process the two process branches (edge
deletion and edge addition) can be analyzed independently and their
contributions can be added together. Similarly as for the SG-ESP process, the
probability that vertex $V_i$ is of degree $k_i$, vertex $V_j$ is of degree
$k_j$ and that the conditions mentioned in the step~\ref{VADE:step:insert-loop}
of VADE are not met is given by

\begin{equation} \label{eq:A} A(k_i,k_j) = \frac{f_1(k_i)  f_2(k_j)
N(k_i) \Bigl( N(k_j) - \delta_{k_i,k_j}\Bigr) \Bigl(1-P_{k_i,k_j}
\Bigr)}{N^{2} \langle f_1 \rangle \langle f_2 \rangle} \end{equation}

The probability that vertex $V_l$ is of degree $k_l$, vertex $V_m$ is of degree
$k_m$ and that the conditions mentioned in the step \ref{VADE:step:delete-loop}
of VADE are not met is given by

\begin{equation} \label{eq:D} D(k_l,k_m) = \frac{g_1(k_i)  g_2(k_j)
N(k_m) \Bigl( N(k_m) - \delta_{k_l,k_m}\Bigr) P_{k_l,k_m}}{N^{2} \langle g_1
\rangle \langle g_2 \rangle} \end{equation}

After the analysis of all situations at time $t$ which will contribute to
$L(k,k')$ at time $t+1$ (there is a total of 48 such situations, see
Table~\ref{table:varVADE}), and after considering that $A(k,k')$ as well as
$D(k,k')$ are both symmetric in $k,k'$, the master equation for the edge
distribution can be written as (again, for notational simplicity the time
variable is suppressed on the right hand side)

\begin{widetext}
\begin{eqnarray}
\label{mult:VADE}
L^{t+1}(k,k') = L(k,k') +2q \sum\limits_{X}\Biggl[A(k-1,X)
\left(T(k-1,k')+\frac{1}{2} \delta_{k'-1,X}\right)+A(X,k'-1)
\left(T(k'-1,k)+\frac{1}{2} \delta_{k-1,X} \right)&& \nonumber\\ -
A(k,X)T(k,k') - A(X,k')T(k'k) \Biggr] + 2(1-q) \sum\limits_{X}
\Biggl[D(k+1,X)T(k+1,k')  + D(X,k'+1)T(k'+1,k) && \nonumber\\ -
D(k,X)\left(T(k,k') - \frac{1}{2}\delta_{k',X}\right) - D(X,k') \left( T(k',k)
- \frac{1}{2}\delta_{X,k}\right)  \Biggr].
\end{eqnarray}
\end{widetext}

This equation is valid for $k \neq k'$. If $k=k'$, the summation term on the
right hand side must be multiplied with the factor $1/2$. The four terms
proportional to $q$ account for the case when an edge is added, the four terms
proportional to $1-q$ account for the case when an edge is deleted. The terms
$T(k,k')$ account for the edges that change their degree because they
participate in one of the involved vertices, whereas the terms $\delta_{k,k'}$
account directly for the edge that is added or deleted.

The aim of introducing the VADE process is mainly to show that certain
processes can lie in the complexity hierarchy between processes described on
the level of the degree distribution and processes having very high description
complexity as SG-ESP. VADE also illustrates, which selection rules would bring
the most problems thus providing a direction where the next modeling efforts
could concentrate. The description of VADE is closed on the edge level and
therefore definitively solvable. However, to understand more deeply the
relation between SG-ESP and VADE more research would be needed to solve its
complex and nonlinear master equation.

\begin {figure}
\begin{center}
\includegraphics[width=8.5cm,
keepaspectratio]{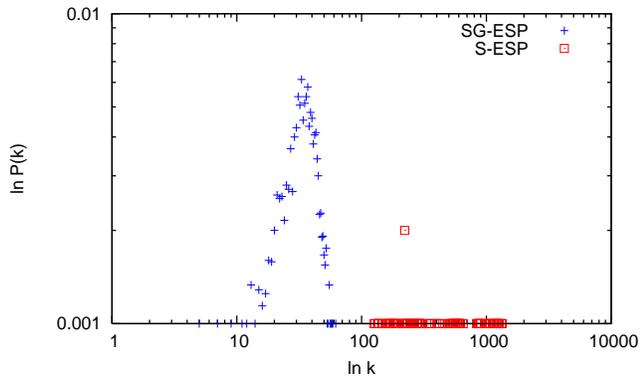}
\end{center}
\caption{\label{fig:experiment-diff} (color online) Simulation experiments for
a network with N=1000, L=2000. The x-axis represents degree values and the
y-axis the values of $P(k)$, both axes are in logarithmic scale. The processes
were simulated over $2.10^6$ iterations to the equilibrium state. The figure shows
the simulation results for the S-ESP and SG-ESP processes with $f(k)=k$. To
obtain the resulting graph a time averaging over 15 instances separated with $10^4$
states was computed. The S-ESP process condensates and all edges are rewired to
a few nodes (finally to only one node) as self-connections.}
\end {figure}

\section{\label{sec:simulations} Simulations}

To numerically support the analysis we described in the previous sections, we
simulated the S-ESP, SG-ESP and VADE processes. Our experiments focus on small
networks and a very long simulation time where a difference between the S-ESP and
SG-ESP processes is clearly visible. In all cases we first generated a
classical random network with Poisson distribution and average degree 4.0,
which was then used as initial condition for all experiments.

\begin {figure}
\begin{center}
\includegraphics[width=8.5cm,
keepaspectratio]{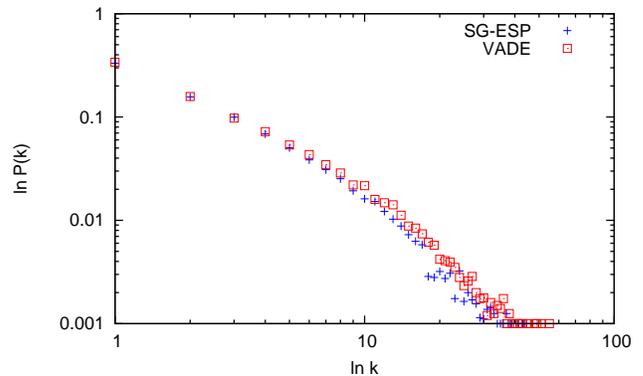}
\end{center}
\caption{\label{fig:experiment-vade} (color online) Simulation experiments
illustrating the similarity between the distributions $P(k)$ of VADE and SG-ESP
processes. The meaning of the axes is the same as in
Figure~\ref{fig:experiment-diff} and the same averaging procedure was used. The
parameters for the SG-ESP process are: N=1000, L=2000 and $f(k)=k$ except for
$f(0)=5$. The network stabilizes in an equilibrium state with a distribution
near to the distribution of a scale-free network. The VADE process was
simulated with the parameters: $N=1000, q=0.004, f_1(k)=k, f_1(0)=1000,
f_2(k)=k, g_1=1, g_2=1$.}
\end {figure}

The experimental results in Figure~\ref{fig:experiment-diff} show that in
situations where the effects of simple graph constraints and of finite network
size cannot be neglected, S-ESP and SG-ESP behave in a very different way.
S-ESP condensates, on the contrary SG-ESP is developing a highly interconnected
kernel. The experimental results for the range of parameters in
Figure~\ref{fig:experiment-diff} also point to the analysis in
\cite{mayerdorog}, especially in relation to the analysis of the dense network
kernel.

Another conclusion from the simulation experiments (see
Fig.~\ref{fig:experiment-vade}) is that the VADE process can approximate the
SG-ESP process very closely due to the flexibility caused by the preferential
parameters $f_1, f_2, g_1,g_2$. This fact opens a possibility of having new
processes which have a closed master equation but which can approximate the
SG-ESP process. Figure~\ref{fig:experiment-vade} shows that the VADE process
generates a scale free network with very small difference to SG-ESP.

In Figure~\ref{fig:experiment-edge} the number of edges for the VADE process is
plotted. The experiments show that if the parameter $q$ of VADE is correctly
set the number of edges will stay stable and it will approximate the SG-ESP
process. The parameter $q$ can be estimated using the following approximation:
the number of edges will stay constant on average if the probability of adding
an edge between two vertices with degree $k,k'$ is equal to the probability of
deleting an edge between two vertices with degree $k,k'$. So one has

\begin{eqnarray} q A(k,k') = (1-q) D(k,k') && \nonumber\\ \Rightarrow q =
\frac{1}{1 + A(k,k')/D(k,k')} && \nonumber\\  = \frac{1}{1 +
\frac{f_1(k)f_2(k')(1-P_{k,k'}) \langle g_1 \rangle \langle g_2 \rangle
}{g_1(k)g_2(k') P_{k,k'} \langle f_1 \rangle \langle f_2 \rangle}}
\end{eqnarray}

Assuming that $f_1(k) = g_1(k)$ and $f_2(k') = g_2(k')$ this equation will
simplify to
\begin{equation*}
q = P_{k,k'}\approx \frac{2L P(k,k')}{N^2P(k)P(k')} \approx
\frac{2L}{N^2}.
\end{equation*}
For the last approximation step an uncorrelated network is assumed and both $k$
and $k'$ are replaced with $\bar{k}$. For the network parameters in Fig.
\ref{fig:experiment-vade}, a value of $q=0.004$ is obtained. This value was
used during the experiments and the results are approximating the objective
value $L=2000$ very closely (see Figure~\ref{fig:experiment-edge}).

\begin {figure}
\begin{center}
\includegraphics[width=8.5cm,
keepaspectratio]{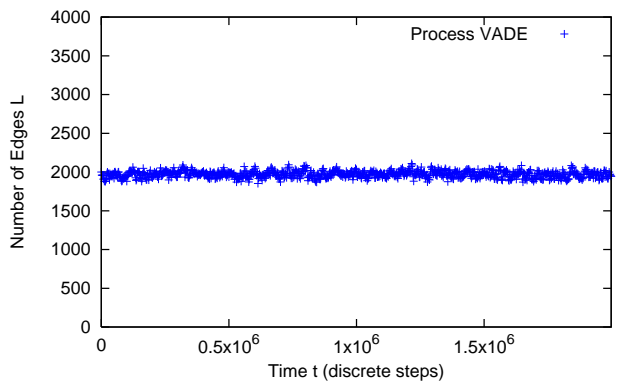}
\end{center}
\caption{\label{fig:experiment-edge} (color online) The figure shows the
simulation results for the VADE process with the same parameters as in
Figure~\ref{fig:experiment-vade}. The x-axis represents the discrete time and
the y-axis the number of edges, both axes are in linear scale. The simulation
experiments show that the number of edges becomes quickly stable.}
\end{figure}

\section{Conclusion}

To summarize, the analysis in Section~\ref{sec:SG-ESP} has shown a method how
to describe the process constraining steps that keep a network evolution
process in the class of simple graphs. To model these constraints we introduced
a new distribution that described wedges - the paths of length 2. The
relation between edge distribution (degree-degree correlation) and wedges was
also studied to understand the evolution of simple graphs. A combination of
such constraints with simple edge selection rules can lead to a very high
complexity involving several higher order distributions. How far in the
distribution hierarchy one has to go to obtain a closed description of the
SG-ESP process is not clear and it would need more research to understand the
situation fully. The study of simple graph constraints provides further reasons
why it is worthwhile studying (and measuring on real networks) the higher order
distributions as for example the wedge distribution.

Up to now, the processes were defined in such a way, that it was possible to
formulate a closed master equation for the degree distribution. However these
processes either did not respect the simple graph structure or were limited in
some other way. When assuming that self-organization is the driving force for
the emergence of complex network structures without self-loops and multiple
connections, higher order distributions are key, in particular for non-growing
simple networks.

The present paper opens several directions for possible research. Further
approximations of equation~\ref{eq:swap} (or process modes and constraints) can
provide deeper insight into the processes on simple graphs or into the
parameter ranges where they can be better understood. Another possibility is to
study in more detail the intermediate VADE class of processes which are more
complex than classical random graphs but simpler than processes containing edge
selection rules. These processes can bring more light into the immense
complexity of simple graphs through approximation of edge selection rules with
more complex set of vertex selection rules.

\begin{acknowledgments}

We would like to thank Professor Peter Widmayer who suggested this interesting
area to us and has continually supported our research, Professor
Narsingh Deo and Professor Joerg Nievergelt for inspiring discussions. The
authors also wish to thank the anonymous referees for their valuable comments
in improving the manuscript, and the linguist Richard Hall for help with
proof-reading.

\end{acknowledgments}

%\newpage %Just because of unusual number of tables stacked at end

\bibliography{../../../lib/references/network}% Produces the bibliography via BibTeX.
}
\bibliographystyle{plain}
\end{document}